\documentclass[fleqn,usenatbib]{mnras}

\usepackage{newtxtext,newtxmath}
\usepackage[T1]{fontenc}

\DeclareRobustCommand{\VAN}[3]{#2}
\let\VANthebibliography\thebibliography
\def\thebibliography{\DeclareRobustCommand{\VAN}[3]{##3}\VANthebibliography}

\usepackage{graphicx}	% Including figure files
\usepackage{amsmath}	% Advanced maths commands
\usepackage{amssymb}	% Extra maths symbols

\title[Radio detection of Swift J1842.5$-$1124]{MeerKAT radio detection of the Galactic black hole candidate Swift J1842.5$-$1124 during its 2020 outburst}

\author[X. Zhang et al.]{
X. Zhang,$^{1,2}$
W. Yu,$^{1}$\thanks{E-mail: wenfei@shao.ac.cn}
S.E. Motta,$^{3,4}$
R. Fender,$^{4,5}$
P. Woudt,$^{5}$
J. C. A. Miller-Jones$^{6}$
and G.R. Sivakoff$^{7}$
\\
$^{1}$Shanghai Astronomical Observatory, Chinese Academy of Sciences, 80 Nandan Road, Shanghai 200030, China\\
$^{2}$University of Chinese Academy of Sciences, 19A Yuquanlu, Beijing 100049, China\\
$^{3}$Istituto Nazionale di Astrofisica, Osservatorio Astronomico di Brera, via E.\,Bianchi 46, 23807 Merate (LC), Italy\\
$^{4}$Department of Physics, University of Oxford, Denys Wilkinson Building, Keble Road, Oxford OX1 3RH, UK\\
$^{5}$Department of Astronomy, University of Cape Town, Private Bag X3, Rondebosch 7701, South Africa\\
$^{6}$International Centre for Radio Astronomy Research - Curtin University, GPO Box U1987, Perth, WA 6845, Australia\\
$^{7}$Department of Physics, University of Alberta, CCIS 4-181, Edmonton, AB T6G 2E1, Canada\\
}

\date{Accepted XXX. Received YYY; in original form ZZZ}

\pubyear{2021}

\begin{document}
\label{firstpage}
\pagerange{\pageref{firstpage}--\pageref{lastpage}}
\maketitle

\begin{abstract}
Swift J1842.5$-$1124 is a transient Galactic black hole X-ray binary candidate, which underwent a new outburst in May 2020. We performed multi-epoch MeerKAT radio observations under the ThunderKAT Large Survey Programme, coordinated with quasi-simultaneous Swift/XRT X-ray observations during the outburst, which lasted nearly a month. We were able to make the first-ever radio detection of this black hole binary with the highest flux density of 229$\pm$31 $\mu$Jy when the source was in the hard state, after non-detection in the radio band in the soft state which occurred immediately after its emergence during the new X-ray outburst. Therefore, its radio and X-ray properties are consistent with the disk-jet coupling picture established in other black hole X-ray binaries. We place the source's quasi-simultaneous X-ray and radio measurements on the radio/X-ray luminosity correlation plane; two quasi-simultaneous radio/X-ray measurements separated by 11 days were obtained, which span $\sim$ 2 dex in the X-ray luminosity. If the source follows the black hole track in the radio/X-ray correlation plane during the outburst, it would lie at a distance beyond $\sim$ 5 kpc.
%as, if we take the lowest detected luminosity as its quiescent X-ray luminosity at below 10$^{34}$ ergs\,s$^{-1}$. 

\end{abstract}


%%%%%%%%%%%%%%%%%%%%%%%%%%%%%%%%%%%%%%%%%%%%%%%%%%

%%%%%%%%%%%%%%%%% BODY OF PAPER %%%%%%%%%%%%%%%%%%

\section{Introduction}
X-ray binaries (XRBs) are stellar binary systems that consist of a primary star, normally a black hole or a neutron star, and a companion star which feeds material to the former. According to the nature of the primary star, XRBs are generally grouped into black hole X-ray binaries (BH XRBs) and neutron star X-ray binaries (NS XRBs). Accreting XRBs emit high energy X-rays (perhaps up to gamma-rays) all the way down to long wavelength radio radiation. The powerful X-ray emission is thought to be emitted from the accretion disk \citep{1973A&A_Shakura}, therefore becoming a primary probe of the accretion process, while the radio emission traces the jets \citep{2002LNP_Fender,2006csxs_Fender}.

Most Galactic BH XRBs are transient systems, spending most of their life time in quiescence in which they accrete material at very low X-ray luminosity (e.g. $\sim$ 10$^{30}$ $-$ 10$^{33}$ ergs\,{s}$^{-1}$ in the range of 0.3$-$7 keV, see \citealt{Kong_2002}). The systems occasionally evolve to an outburst \citep{1997ApJ_Chen,Yan_2015} phase during which the peak X-ray luminosity can reach (or even exceed, although rarely) the Eddington limit \citep{1997ApJ_Chen}. During an outburst they can span over several spectral and timing states \citep[e.g.,][]{2001ApJS_Homan,Belloni_2010}, each of which exhibits specific X-ray spectral and timing properties \citep[e.g.,][]{Remillard_2006,Belloni_2016} and jet behavior. Generally, at the beginning of an outburst, a transient Galactic BH XRB resides in a so-called hard state wherein the X-ray emission is attributed to the comptonization of seed soft X-ray photons from a radiatively inefficient accretion flow and dominated by power-law radiation. In this stage, there exists a steady, partially self-absorbed compact jet \citep[e.g.,][]{1979ApJ_Blandford,2000ApJ_Dhawan,Fender_2001,2001MNRAS_Stirling}, which exhibits a flat or slightly inverted spectrum ($\alpha \gtrsim 0$ in the form of $S_{\nu} \propto \nu^{\alpha}$, where $S_{\nu}$ is the observed flux density, which scales with frequency, $\nu$; e.g.\ \citealt{2000MNRAS_Fender,Fender_2001}) in the range from radio to IR wavelengths. As the X-ray flux increases, the X-ray spectrum become progressively dominated by the accretion disc thermal emission. In this phase such a source will transition from the hard to the soft state, moving through the hard-intermediate and soft-intermediate state. At the transition to the soft-intermediate state, relativistic (sometimes apparently super-luminal) transient jets with optically-thin radio spectra can be observed \citep[e.g.,][]{1994_Mirabel_Nature,1999_Fender,2012MNRAS_Miller-Jones,Russell_2019,Bright_2020}, which have been associated with radio flares \citep{1995_Hjellming_Nature,1999_Fender,Bright_2020}. When such a system reaches the soft state, the X-ray emission is thermal, and the compact jet we can see in the hard state is thought to be quenched. As the mass accretion rate drops and the disk cools, the outburst begins to fade. The BH XRB then transitions back to the hard state through the intermediate states, finally returning to quiescence. A typical outburst is usually expected to last from weeks to months, but rarely even years \citep[e.g.,][]{1997ApJ_Chen,Yan_2015}.

During the hard state, the radio emission from a BH XRB is correlated to its X-ray emission \citep{1998A&A_Hannikainen,2000A&A_Corbel,2003A&A_Corbel,Gallo_2003}, %which generates a correlation in the form of $L_R\propto L_X^\alpha$ ($\alpha$ typically around 0.6, $L_R$ and $L_X$ are measured luminosities), 
and this correlation extends down to quiescence \citep{2013MNRAS_Corbel,Gallo_2014,2020MNRAS_Tremou}. Studies of BH XRBs have highlighted the so-called `standard' (or radio-loud) track with radio and X-ray luminosities taking the relation of $L_R\propto L_X^{\sim 0.6}$ \citep{2013MNRAS_Corbel}, as well as another track which is sometimes referred to as `radio-quiet' \citep[e.g.,][]{2004ApJ_Corbel,Rodriguez_2007}, and has a different correlation of $L_R\propto L_X^{\sim 1.4}$ \citep{2013MNRAS_Corbel}. Notably, some BHs can switch from a high X-ray luminosity radio-quiet track to a low X-ray luminosity radio-loud track \citep[e.g.,][]{Coriat_2011}. There are a number of scenarios proposed to explain the observed two tracks, including, for example, differences in the jet magnetic field \citep{Casella_2009}, accretion flow radiative efficiency \citep{Coriat_2011}, comptonization of additional photons from an inner disk \citep{Meyer_Hofmeister_2014}, and inclination angle \citep{Motta_2018}. This radio/X-ray correlation has became an important piece of observational evidence supporting the connection between accretion processes and the production of jets (i.e.\ the disk-jet coupling, \citealt{Fender_2004}; also holds for NS XRBs, e.g.\ \citealt{Migliari_2006}). Compared to AGNs and GRBs, which evolve on much longer and much shorter time-scales respectively, BH XRBs are ideal laboratories in which to probe the disk-jet coupling, because they show spectra and fast time variability on humanly accessible time-scales (from years down to sub-second time-scales).  
%BH XRBs are ideal objects to disentangle the poorly understood mechanism of how disk-jet are coupled and how jets are formed considering its proper timescales of weeks to years for a typical quiescence-outburst-quiescence cycle, in comparison to objects that harbour jets like Active Galactic Nucleus (AGNs) and Gamma ray bursts (GRBs) with much longer and shorter timescales, respectively. 
\smallskip 

\subsection{Swift J1842.5--1124}
\label{sec:1_1} % used for referring to this section from elsewhere
Swift J1842.5$-$1124 is a Galactic plane source, which was first detected in the X-ray band during its 2008 outburst \citep{2008ATel.1610....1K}. This source had been covered by a number of follow-up multi-wavelength observations, leading to further detections in X-ray \citep[]{2008ATel.1706....1K,2008ATel.1716....1M}, ultraviolet/optical \citep{2008ATel.1716....1M} and near-infrared \citep{2008ATel.1720....1T} bands. \citet{2008ATel.1716....1M} suggested that the source is a BH binary that was evolving in the hard state, according to the X-ray spectral properties (e.g., combined black body and power law model with black body kT of 0.9 keV and photon index of 1.5) and the detection of a strong QPO at $\approx$ 0.8 Hz, together with the strong X-ray variability. In addition, the observed timing properties which were systematically studied by \citet{Zhao_2016}, are similar to other BH XRBs in many respects, supporting a BH nature of this source. Furthermore, the peak of the hard X-ray light curve precedes the peak of the softer X-ray light curves by $\sim$ 10 days \citep{Krimm_2013}, which is larger than what has been seen in neutron star LMXB transients \citep{2003ApJ...589L..33Y,2004ApJ...611L.121Y}, supporting the scenario in which the source is a candidate BH XRB \citep[e.g.,][]{Krimm_2011,Brocksopp_2005}.

On May 23, 2020 (MJD 58992), a new outburst from Swift J1842.5$-$1124 was detected \citep{2020ATel13762....1S} by Monitor of All-sky X-ray Image/Gas Slit Camera (hereafter \textit{MAXI}; \citealt{2009PASJ_Matsuoka}) with a daily averaged 2--10 keV intensity of $\sim$ 13 $\pm$ 3 mCrab. Then it increased in X-ray flux, reaching an intensity of 17 $\pm$ 3 mCrab on March 25 (MJD 58994), confirming the source had entered into a new outburst. We started to monitor the source with the Meer Karoo Array Telescope (MeerKAT; \citealt{2009IEEEP_Jonas}) as part of the ThunderKAT\footnote{The HUNt for Dynamic and Explosive Radio transients with meerKAT: \url{http://www.thunderkat.uct.ac.za}} Large Survey Programme \citep{fender2017thunderkat} on a weekly basis from June 1 (MJD 59001) until June 27 (MJD 59027) for about one month and radio detection has been reported based on preliminary analysis \citep[]{2020ATel_xian_a,2020ATel_xian_b}. Accompanying X-ray observations include observations with the X-ray Telescope (XRT) instrument on board the Neil Gehrels Swift Observatory (\textit{Swift}; \citealt{2004ApJ_Gehrels}) and daily monitoring data from the Burst Alert Telescope (BAT) on board Swift, as well as daily monitoring data from MAXI.

\section{Observations and data reduction}
\subsection{MeerKAT observations and data reduction}

We conducted near-weekly MeerKAT pointed observations of Swift J1842.5$-$1124 for five consecutive weeks under the X-ray binary monitoring programme in ThunderKAT. The details of the observations are listed in Table~\ref{table:1}. Each observation performed has $\sim$ 15 minutes of on-source exposure time. Data were taken in L band (centered at 1.284 GHz) with a bandwidth of 856 MHz, which was originally split in 32768 channels, and which we re-binned to 512 channels before data reduction. J1939-6342 was used as the bandpass and flux density calibrator to set the flux scale and solve the bandpass, and J1833-2103 was used as phase calibrator to determine the complex gains as a function of time. Flagging and calibration were done using the Common Astronomy Software Application package (CASA 4.7.1, hereafter CASA; \citealt{2007ASPC_McMullin}), applying standard procedures\footnote{\url{https://casaguides.nrao.edu/index.php?title=Karl_G._Jansky_VLA_Tutorials}}. After that, we did another round of flagging based on each calibrated measurement set, to excise bad data that still remained. For this we used the task FLAGDATA, using rflag and extend mode along the time and frequency axes, respectively. Finally, we split off the target field, and imaged it in total intensity (Stokes I) using WSCLEAN \citep{2014MNRAS_Offringa}. We used a Briggs weighting scheme with a robust parameter of $-0.7$ to suppress the side-lobe effects across the field. Besides, the auto-thresholding algorithm was invoked for the clean. We made use of the CASA task IMFIT to measure the flux density of the source for all epochs by fitting a point source in the image plane with an elliptical Gaussian model fixed to the size and shape of the restoring beam. The local rms noise in each observation was obtained from a close-by source-free region. Details of the measured flux densities and rms noises are shown in Table~\ref{table:1}.

\subsection{X-ray observations and data reduction}
\subsubsection{Swift/XRT observations}
\label{sec:2_2_1}
Throughout the 2020 outburst of Swift J1842.5$-$1124, the source was targeted by the XRT instrument on board the Swift Observatory with a cadence of about a week. We extracted one energy spectrum per observation using the UK Swift Science Data Centre pipeline, which provides publication-quality data \citep{2007A&A_Evans,2009MNRAS_Evans}. There are 8 Swift/XRT observations covering the near-one-month outburst of the source, but only three of them resulted in a detection. We fitted each spectrum  within XSPEC \citep{1996ASPC_Arnaud} with either a powerlaw model (powerlaw in XSPEC) or a disk blackbody (diskbb in XSPEC) model,  modified by interstellar absorption. We took the average of the fitted Galactic neutral hydrogen absorption column density in the two epochs (MJD 59000 and 59010) after they were freely varied in individual fitting when the source was significantly detected (individual best-fit values are $0.387^{+0.008}_{-0.008}$, $0.248^{+0.028}_{-0.027}$ $\times{10^{22}\,{\rm  cm}^{-2}}$, errors are quoted as 1 $\sigma$). Then we froze to the averaged value, which is $\rm N_H=0.318\times{10^{22}\,{\rm  cm}^{-2}}$, across all three spectrum files. We found this value well-matches the Galactic value in the direction of the source ($\rm N_H=0.317\times{10^{22}\,{\rm  cm}^{-2}}$; \citealt{2016A&A_HI4PI}). The unabsorbed fluxes in the 1--10 keV band were extracted with the task CFLUX. The results of the fitting are listed in Table~\ref{table:2}. 

\subsubsection{MAXI and Swift/BAT monitoring observations}
\label{sec:2_2_2}
The daily and orbital X-ray light curves of the source are publicly available\footnote{\url{http://maxi.riken.jp/top/slist.html}}. We extracted the source's daily-averaged light curve in the soft X-ray band obtained with MAXI. Similarly, we also extracted the 15--50 keV light curve obtained with Swift/BAT\footnote{\url{https://swift.gsfc.nasa.gov/results/transients/BlackHoles.html}}.

\begin{figure}
	\includegraphics[width=\columnwidth]{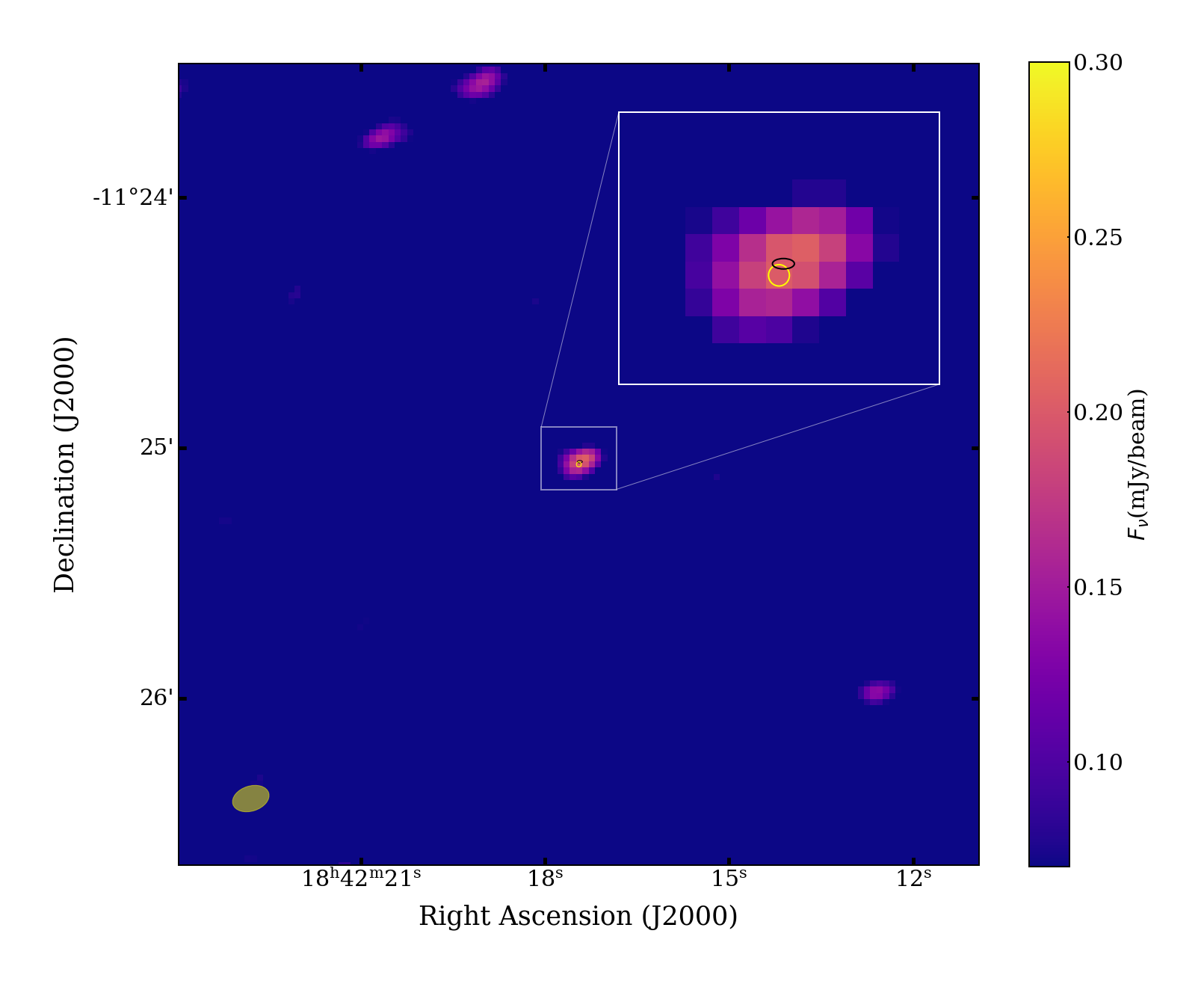}
    \caption{The MeerKAT zoomed-in radio image of Swift J1842.5$-$1124 of epoch 2 at 1.284 GHz. %The source lies at the center of the image, whose position is consistent with other reported multi-wavelength counterparts
    The radio position of the source, marked by the black ellipse, is consistent with its ultraviolet/optical position \citep{2008ATel.1716....1M}, marked by the yellow circle, see Section~\ref{sec:3_1} for details. The synthesized beam is marked with a yellow ellipse with a size of $9.19^{\prime \prime}$$\times$$6.01^{\prime \prime}$ and shown at the bottom left of the image.}
    \label{fig:example_figure1}
\end{figure}

\begin{figure}
	
	\includegraphics[width=\columnwidth]{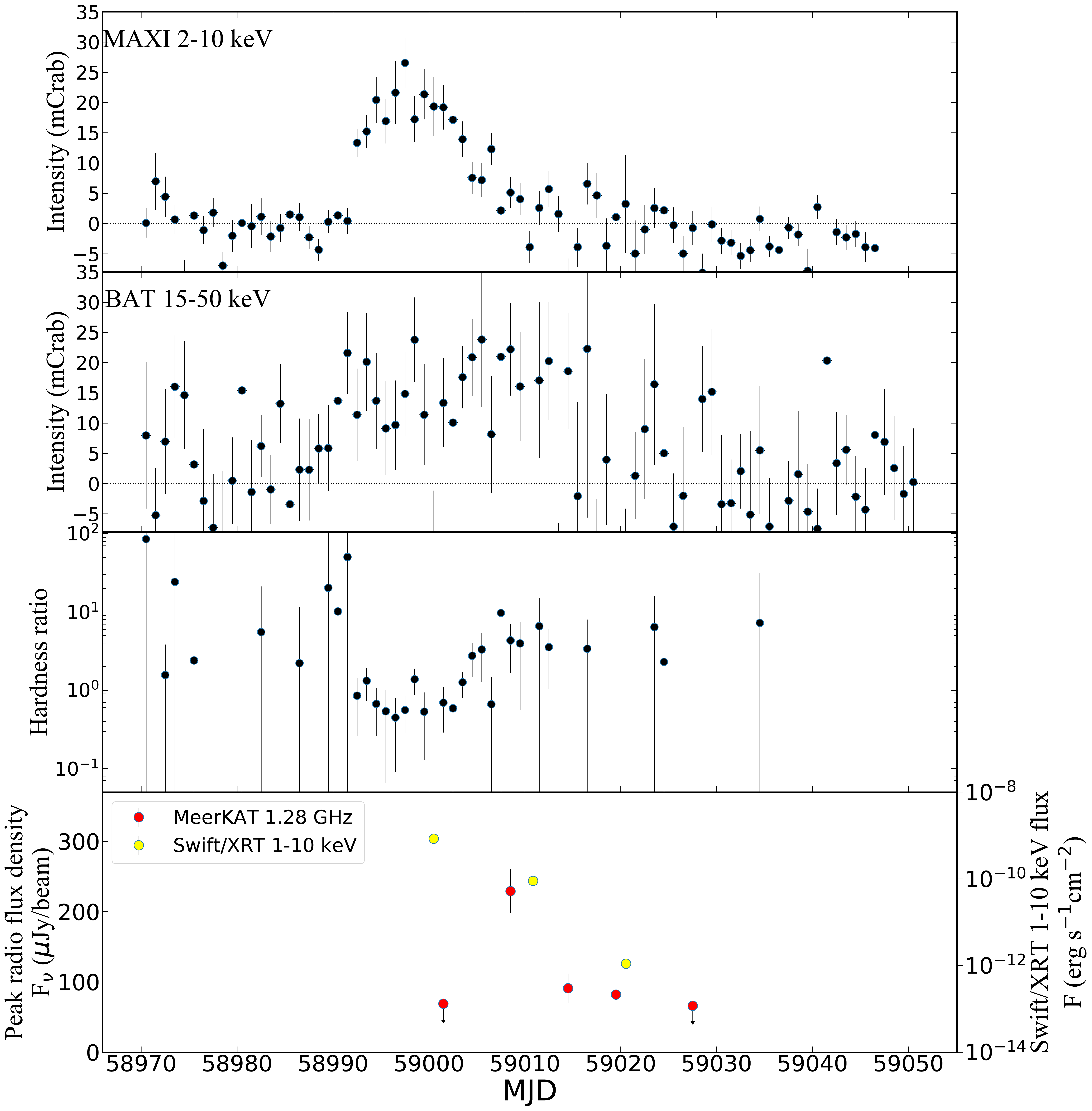}
    \caption{X-ray and radio light curves of Swift J1842.5$-$1124 as seen with MAXI, Swift and MeerKAT. \textit{First panel}: MAXI 2$-$10 keV light curve of Swift J1842.5$-$1124 throughout the 2020 outburst. \textit{Second panel}: BAT 15$-$50 keV light curve of Swift J1842.5$-$1124 throughout the 2020 outburst. \textit{Third panel}: Hardness ratio calculated as BAT 15$-$50 keV intensity over MAXI 2-10 keV intensity on one-day-averaged basis. \textit{Fourth panel}: MeerKAT radio (red circles) and Swift/XRT (yellow circles) detections of Swift J1842.5$-$1124.}
    \label{fig:example_figure2}
\end{figure}

\begin{table*}
\small
\caption{Summary of 1.28 GHz MeerKAT radio observations of Swift J1842.5$-$1124.}\quad

\begin{tabular}{ccccccc}
\hline\hline
{\bf Epoch}&{\bf Date} &{\bf MJD$^a$}&{\bf RMS noise$^b$}&{ \bf Flux density$^c$}&{\bf Spectral index$^d$}&{\bf Spectral state}\\
&{\bf (2020)}&{\bf(J2000)}&{\bf($\mathbf{\mu {\rm \bf Jy}\,{\rm \bf beam}^{-1}}$)}&{\bf($\mathbf{\mu {\rm \bf Jy}}$)}& \\[0.2cm]
\hline
\phantom{0}1$^e$&June 1&59001.124&23&$<69$&--&Soft state\\[0.15cm]
\phantom{0}2$^f$&June 8&59008.171&26&229$\pm$31&0.87$\pm$0.59&Hard state\\[0.15cm]
3&\phantom{0}June 14&59014.892&24&\phantom{0}91$\pm$21&-1.89$\pm$1.12&Hard state\\[0.15cm]
\phantom{0}4$^f$&\phantom{0}June 19&59019.058&23&\phantom{0}82$\pm$18&--&Hard state\\[0.15cm]
\phantom{0}5$^e$&\phantom{0}June 27&59027.008&22&$<66$&--&--\phantom{0}$^g$\\[0.20cm] \hline
\end{tabular}\\
\begin{flushleft}
{$^a$ All mark the start point of the observations, each lasts 15 minutes.}\\
{$^b$ RMS noise is measured from a close-by source free region.}\\
{$^c$ Radio flux density, where uncertainties are quoted at the $1\sigma$ level.}\\
{$^d$ Measured in-band spectral indices using the 856 MHz bandwidth centered at 1.28 GHz, see Section~\ref{sec:3_2} for details. These values should be treated with caution due to the limited bandwidth and low significance of the detection.} \\
{$^e$ The source was not detected in both epochs. $3\sigma$ upper limits are given.}\\
{$^f$ Both epochs have paired XRT observation within 3 days to study the radio/X-ray correlation.}\\
{$^g$ The spectral state can not be determined for this observation.The source might have entered the quiescent state, see Section~\ref{sec:3_3}.}
\end{flushleft}

\label{table:1}
\end{table*}

\begin{table*}
\small
\caption{Summary of the Swift XRT observations with detections.}\quad
\centering
\begin{tabular}{ cccccccccc }
  \hline\hline
  {\bf Obs. ID}&{\bf Date}&{\bf MJD}$^a$&{\bf Mode}&{\bf Exposure time}&{\bf $\mathbf{N_H}$}$^b$&{\bf $\mathbf{\Gamma}$}$^c$&{\bf kT}$^d$&{\bf $\mathbf{F_{unabs}}$}$^e$&{\bf $\mathbf{\chi^2}$}\\
  &{\bf (2020)}&{\bf (J2000)}&&{\bf (s)}&{\bf $\mathbf{(10^{22}\,{cm}^{-2})}$}&&{\bf (keV)}&{\bf ($\mathbf{10^{-10}{erg\,s}^{-1}{cm}^{-2}}$)}&\\[0.2cm] 
  \hline 
00031234016&May 31&59000.496&WT&831.564&$0.318$&--&$0.573^{+0.002}_{-0.002}$&\phantom{0}$8.404^{+0.036}_{-0.036}$&1.67\\[0.15cm]
\phantom{0}00031234017$^f$&June 10&59010.848&WT&1179.559&0.318&\phantom{0}\phantom{0}$1.827^{+0.027}_{-0.026}$&--&\phantom{0}$0.894^{+0.018}_{-0.018}$&1.03\\[0.15cm]
\phantom{0}00031234018$^f$&June 20&59020.545&WT&824.088&0.318&\phantom{0}\phantom{0}$2.343^{+2.131}_{-0.818}$&--&\phantom{0}$0.011^{+0.011}_{-0.004}$&1.29\\[0.20cm]\hline

\end{tabular}\\
\begin{flushleft}
{$^a$ All mark the start point of the observations.}\\
{$^b$ $\rm N_H$ represents the fixed Galactic neutral hydrogen absorption column density used across spectra, see Section~\ref{sec:2_2_1} for details. }\\
{$^c$ $\Gamma$ is the fitted power-law photon index. }\\
{$^d$ T shows the fitted DISKBB temperature.}\\
{$^e$ 1--10 keV unabsorbed flux, uncertainties are quoted at the $1\sigma$ level.}\\
{$^f$ The source was in the hard state, when there were paired MeerKAT radio observations (epoch 2 and 4 respectively) to study the radio/X-ray correlation.}\\
\end{flushleft}
\label{table:2}
\end{table*}
\renewcommand\tabcolsep{6.0pt}

\section{Results}
\subsection{Radio detections and position}
\label{sec:3_1}
We detected Swift J1842.5$-$1124 in three of the five MeerKAT observations we obtained. In all three observations the target appears as an unresolved source at the phase center. The target was detected at high statistical significance ($\sim$ 8.8 $\sigma$) in epoch 2, and with marginal significance ($\sim$ 3.8 $\sigma$ and $\sim$ 3.6 $\sigma$) in epochs 3 and 4. We attempted to stack the observations of these three epochs in the uv-plane to try to get a well-refined radio position. However, the radio position returned from the CASA IMFIT task has larger uncertainties than that obtained with data of epoch 2, i.e., the corresponding refined position given by IMFIT for the epoch 2 data is:
\begin{eqnarray}
\nonumber 
{\rm RA \rm (J2000):}&\,18^{\rm h}42^{\rm m}17.433^{\rm s}\pm 0.042^{\rm s} \\ \nonumber
{\rm DEC \rm (J2000):}&\,-11^{\circ}25\arcmin03.263\arcsec\pm 0.287\arcsec \nonumber
\end{eqnarray}
%Where the second terms above are the IMFIT task fitted statistical errors. 
The above radio position and positions measured from epoch 3, 4 data are consistent within 1 $\sigma$. In Fig.~\ref{fig:example_figure1}, we present the radio image of the detection of Swift J1842.5$-$1124 (located at the center of the image) obtained in epoch 2, when the source was brightest. The radio position is consistent with the published ultraviolet/optical position (\citealt{2008ATel.1716....1M}; estimated 90\% confidence uncertainty radius of 0.6 arcsec) 0.69 arcseconds away. In addition, the radio position is within 3.08 arcseconds of the reported X-ray position (\citealt{2008ATel.1610....1K}; estimated error of 1.8 arcsec radius), and is within 0.69 arcseconds of the position of the near-infrared counterpart obtained with the Baade telescope \citep{2008ATel.1720....1T}. Combined with the variability of the radio source, we conclude that this is the radio counterpart of Swift J1842.5$-$1124.

Throughout the 2020 outburst, no distinct transient jet ejecta were observed from Swift J1842.5$-$1124 in our MeerKAT monitoring observations.

\subsection{Radio spectrum}
\label{sec:3_2}
In order to investigate the radio spectrum of Swift J1842.5$-$1124 during the MeerKAT observation campaign, we split the total 856 MHz bandwidth into two parts, each of width 428 MHz (centered at 1.07 GHz and 1.5 GHz, respectively) for the epochs in which we detected the source. We then imaged the source field in the two sub-bands with WSCLEAN and measured the corresponding flux density of the source. Then we were able to calculate the spectral index based on the formalism of $S_\nu \propto \nu^\alpha$. The spectral index uncertainties were calculated with the formula:
\begin{equation}
\delta(\alpha) = \frac{1}{\ln (\frac{\nu_{1}}{\nu_{2}})} \sqrt{\left(\frac{\delta(S_{\nu_{1}})}{S_{\nu_{1}}}\right)^{2}+\left(\frac{\delta(S_{\nu_{2}})}{S_{\nu_{2}}}\right)^{2}} 
\label{eq:equation_1}
\end{equation}
where $\alpha$ is the spectral index, $\nu_{1}$ and $\nu_{2}$ are the observing frequencies, at which $S_{\nu_{1}}$, $S_{\nu_{2}}$ are the measured flux densities. All spectral index measurements are shown in Table~\ref{table:1}.

On June 8, 2020 (MJD 59008), Swift J1842.5$-$1124 was significantly detected by MeerKAT, which resulted in a spectral in-band index of 0.87 $\pm$ 0.59. This value is slightly higher than the typical spectral index observed in other BH XRBs in the hard state \citep{2018MNRAS_Espinasse}, but still consistent with a flat or slightly inverted spectrum. Then on June 14 (MJD 59014), we measured a spectral index of -1.89 $\pm$ 1.12, which is consistent with a steep, optically thin spectrum. The large uncertainties are ascribed to the narrow frequency range we used to do the calculation as well as the low significance of the flux density we observed. If correct, the above values suggest a switch from optically thick radio emission, to (partially) self-absorbed optically thin synchrotron emission.

\subsection{Evolution of the radio and X-ray emission}
\label{sec:3_3}
In Fig.~\ref{fig:example_figure2}, we show the MAXI (2-10 keV) and BAT (15-50 keV) daily-averaged light curves. We also put five MeerKAT radio and three XRT X-ray flux measurements together. We made use of the ratio between 15-50 keV from Swift/BAT and 2-10 keV from MAXI to calculate the hardness ratio (HR), as shown in the third panel of this figure. On MJD 58992, Swift J1842.5$-$1124 was found in its 2020 outburst with the intensity of $\sim$ 13 mCrab. Then the source increased in soft X-ray flux, peaked at $\sim$ 27 mCrab on MJD 58997, after which the source's soft X-ray flux was dropping before MJD 59004. During this period (MJD 58992 to 59003) the source was in the soft state, as evidenced by the rather lower HR value ($\lesssim$ 1) and the XRT fitted result showing a diskbb component with a modified inner disk temperature of $\sim$ 0.57 keV on MJD 59000. This is consistent with the quasi-simultaneous MeerKAT non-detection of core jet emission taken on MJD 59001. From MJD 59004 the source kept the same flux for two days and the HR values were both larger than 2, which may indicate the soft-to-hard state transition. On MJD 59006, the source re-brightened in soft X-ray flux and seemed to make an excursion to the soft state (HR < 1). From MJD 59007 on, the source was stably keeping an HR value above 2. Combined with the three MeerKAT radio detections and XRT X-ray power-law spectra in the initial two weeks of the stage, we suggest the source was settling in the hard state. It was then heading towards quiescence. On MJD 59027, the source was not detected by MeerKAT, while the XRT data taken in WT mode does not allow us to derive its X-ray flux confidently. Radio  non-detection might signal the source had returned back to the quiescent state.

\subsection{Radio -- X-ray correlation}
\label{sec:3_4}
%The correlation between the radio and X-ray fluxes for BH was first revealed for GX 339$-$4 \citep{1998A&A_Hannikainen} in its hard state. In soft X-ray band, the relation follows the form of $S_R\propto S_X^{\sim 0.7}$ where $S_R$ and $S_X$ represent measured fluxes \citep{2003A&A_Corbel}. The relation, when taking a scaled distance of 1 kpc and absorption correction (thus measured fluxes are proportional to luminosities) into account, was found to hold for BH XRB V404 Cygni in quiescence as well as a well-represented large sample of hard state BHs \citep{Gallo_2003}. Studies of BH XRBs followed this work have highlighted the so-called 'standard' (or radio-loud) track with radio and X-ray luminosities taking the relation of $L_R\propto L_X^{\sim 0.6}$ \citep{2013MNRAS_Corbel}, as well as another track which is sometimes referred to as radio-quiet (as opposed to the radio-loud track found at higher radio fluxes) \citep[e.g.,][]{2004ApJ_Corbel,Rodriguez_2007}, has the universal relation of $L_R\propto L_X^{\sim 1.4}$ \citep{2013MNRAS_Corbel}. Notably, some BHs can switch from high X-ray luminosity radio-quiet track to low X-ray luminosity radio-loud track \citep[e.g.,][]{Coriat_2011}. There are a number of scenarios proposed to explain the observed two tracks, including, for example, differences in the jet magnetic field \citep{Casella_2009}, accretion flow radiative efficiency \citep{Coriat_2011}, comptonization of additional photons from an inner disk \citep{Meyer_Hofmeister_2014}, inclination angle \citep{Motta_2018}.

\begin{figure}
\centering
    \includegraphics[width=\columnwidth]{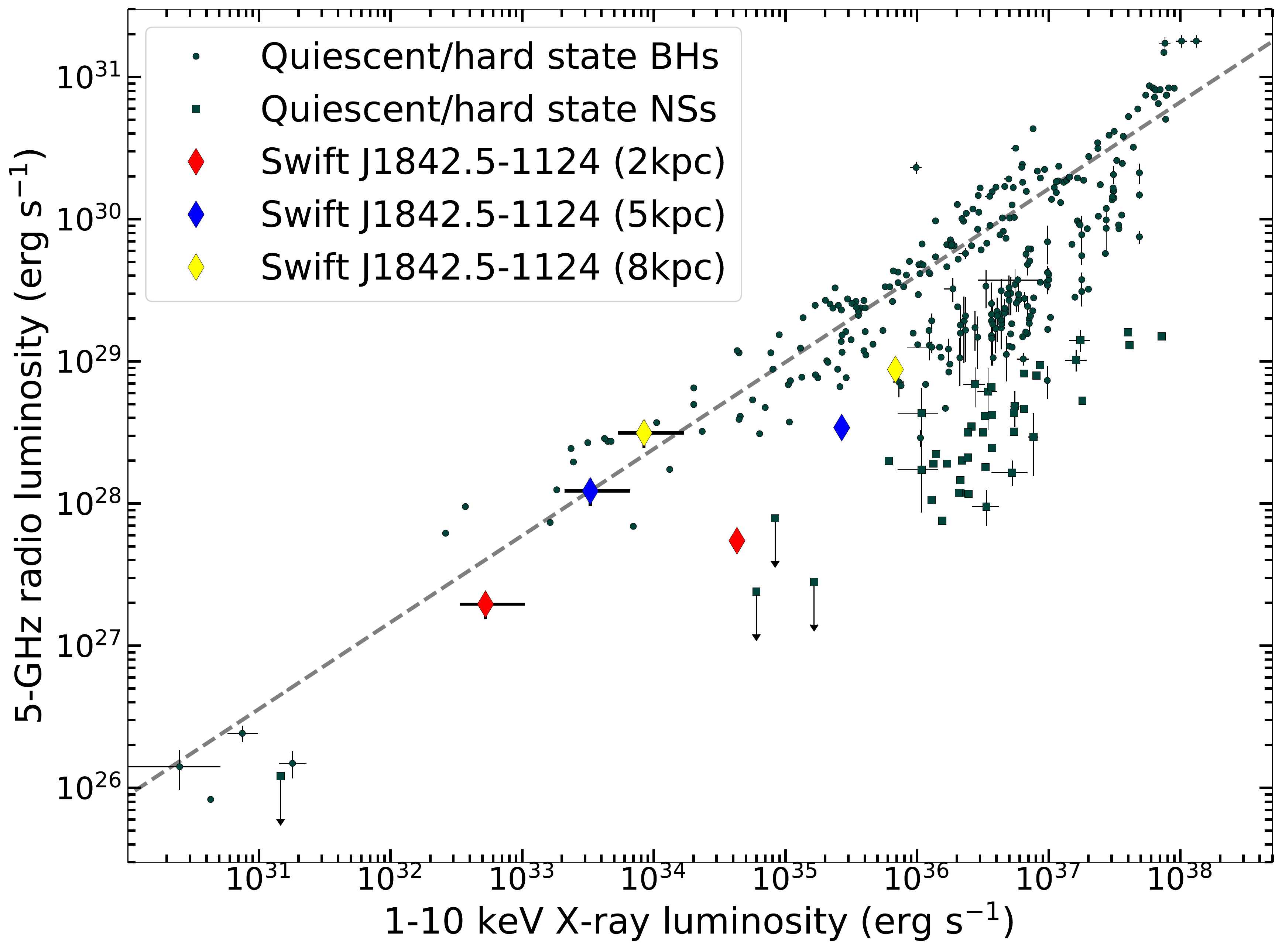}
    \caption{The plot of the radio and X-ray correlation for hard state and quiescent BH XRBs and NS XRBs (denoted by black circles and squares) provided in \citet{Arashbahramian_2018}, together with the hard state measurements of the black hole transient Swift J1842.5$-$1124 assuming a source distances of 2 kpc, 5 kpc, and 8 kpc, respectively, which are marked as red, blue, and yellow diamonds accordingly. %In addition, two blue parallel lines show the luminosities at various distances from 2 - 8 kpc. 
    The dashed line marks the fit for radio-loud track quiescent/hard state BHs collected in \citet{Arashbahramian_2018}, with a correlation of $L_R\propto L_X^{0.61}$. At distances $\gtrsim$ 5 kpc Swift J1842.5$-$1124 appears to join the BH sample.}
    \label{fig:example_figure3}
\end{figure}

In Fig.~\ref{fig:example_figure3}, we plot the quasi-simultaneous radio and X-ray hard state measurements of Swift J1842.5$-$1124 together with the radio/X-ray database collected by \citet{Arashbahramian_2018}. Of all the three MeerKAT detections, we use two pairs of radio and X-ray measurements that were taken within 3 days (see Table~\ref{table:1},\ref{table:2}) to study the radio/X-ray correlation. We excluded the radio detection on MJD 59014 since the source had no quasi-simultaneous X-ray observation. We convert the 1.284 GHz radio flux densities to 5 GHz luminosities by assuming a flat spectrum, while taking the source distance as 2, 5, or 8 kpc to show potential influence of source distance, as we don't have any distance measurement for the source. The 1–10 keV luminosities were estimated based on the 1-10 keV unabsorbed X-ray fluxes. Despite the fact that only two radio/X-ray pairs are shown, the source's X-ray luminosity spanned over $\sim$ 2 orders of magnitude, which is indicative of a much flatter correlation index ($L_R\propto L_X^{< 0.61}$) in comparison to other BH XRBs. Because our sample of measurements is small, Swift J1842.5$-$1124 may follow either the radio-loud or the radio-quiet BH track, or may be another BH system that switches from the radio quiet to radio loud track in an outburst (Notably, ThunderKAT project has revealed a number of sources showing 'track' switching in a single outburst, e.g.\ \citealt{2020MNRAS_Williams,2021MNRAS_Monageng,carotenuto2021hybrid}). Assuming the nature of the source as a BH binary, it should locate relatively distant ($\gtrsim$ 5 kpc) such that 
%as what's generally held for other BHs \citep{Migliari_2006}, the more radio luminous pair can
it joins the BH sample, as shown in Fig.~\ref{fig:example_figure3}.
%As BHs accrete material roughly below the X-ray luminosity of 10$^{34}$ ergs\,s$^{-1}$ \citep[e.g.,][]{2013ApJ_Plotkin} in the quiescent state, we suggest the source, which was probably approaching the quiescent state given its faint X-ray and radio emission seen around MJD 59019-59020, should be at a distance of $\sim$ 8 kpc or more as such a distance would correspond to X-ray luminosity of $\sim$ 10$^{34}$ ergs\,s$^{-1}$ for the less luminous radio/X-ray pair as shown in Fig.~\ref{fig:example_figure3}.

\section{Conclusions}
In this paper, we present our results from the observations of the transient black hole X-ray binary candidate Swift J1842.5$-$1124, which were carried out with the MeerKAT radio observations as part of ThunderKAT and quasi-simultaneous X-ray observations that were obtained from Swift XRT and BAT on board the Swift observatory and the MAXI. We made the first-ever detection of the radio jet with MeerKAT after it transited to the hard state from the soft state when it showed jet quenching during the 2020 outburst, therefore during which the source exhibited the disk-jet coupling showed in other BH XRBs. The quasi-simultaneous MeerKAT radio and Swift/XRT X-ray observations helped investigate the radio/X-ray correlation for the source. In conclusion, our measurements do not follow either BH track in the radio/X-ray correlation plane during the 2020 outburst of Swift J1842.5$-$1124, although the source may cover X-ray luminosity range of about 2 dex and have flat correlation index (< 0.61). Furthermore, the distance of the source would be at about 5 kpc or more, if taking it as a BH binary.
%Furthermore, the distance of the source may be at about 8 kpc or more, if the source's quiescent X-ray luminosity is consistent with those of other black hole binaries.

\section*{Acknowledgements}
XZ would like to acknowledge Dr. Ian Heywood from University of Oxford for critical suggestions on radio data processing and Dr. Payaswini Saikia from New York University Abu Dhabi for useful comments. This work was parlty supported by the National Program on Key Research and Development Project (Grant No. 2016YFA0400804) and the National Natural Science Foundation of China (grant number 11333005 and U1838203). We thank the staff at the South African Radio Astronomy Observatory (SARAO) for their rapid scheduling of these MeerKAT radio observations. The MeerKAT telescope is operated by the South African Radio Astronomy Observatory, which is a facility of the National Research Foundation, an agency of the Department of Science and Innovation. We acknowledge the UK Swift Science Data Centre for building Swift/XRT products. This research has used public data from the Swift/BAT data archive and publicly available data provided by RIKEN, JAXA, and the MAXI team. We acknowledge the use of the ilifu cloud computing facility - www.ilifu.ac.za, a partnership between the University of Cape Town, the University of the Western Cape, the University of Stellenbosch, Sol Plaatje University, the Cape Peninsula University of Technology and the South African Radio Astronomy Observatory. The ilifu facility is supported by contributions from the Inter-University Institute for Data Intensive Astronomy (IDIA - a partnership between the University of Cape Town, the University of Pretoria and the University of the Western Cape), the Computational Biology division at UCT and the Data Intensive Research Initiative of South Africa (DIRISA).

\section*{Data Availability}
The data used in this work can be shared on reasonable request to the corresponding author.



\bibliographystyle{mnras}
\bibliography{SwiftJ1842} % if your bibtex file is called example.bib

%\begin{thebibliography}{99}
%\end{thebibliography}

%%%%%%%%%%%%%%%%%%%%%%%%%%%%%%%%%%%%%%%%%%%%%%%%%%

% Don't change these lines
\bsp	% typesetting comment
\label{lastpage}
\end{document}